\documentclass[osa,twocolumn,amsmath,amssymb,showpacs]{revtex4}     
\usepackage{graphicx}% Include figure files
\usepackage{dcolumn}% Align table columns on decimal point

\begin{document}

%\special{papersize=8.5in,11in}

%\draft

%\wideabs{

\title{Collision rates in near-resonant optical lattices}

\author{Jyrki Piilo}
%\email{jyrki.piilo@helsinki.fi}
\affiliation{Helsinki Institute of Physics,
PL 64, FIN-00014 Helsingin yliopisto,
Finland}

%\date{\today}

%\maketitle

\begin{abstract}
We present a simple method to calculate the binary collision
rate between atoms in near-resonant optical lattices. The method 
is based on the Monte Carlo wave function
simulations and the collision rate is obtained by monitoring
the quantum flux beyond  the average distance between
the atoms. To illustrate the usefulness of the 
method, we calculate the collision rates for
a wide range of occupation densities and various modulation depths of the lattice.
The method presented here 
combined with the semiclassical calculations 
accounting for intra-well collisions can simplify the
study of the effects of binary collisions 
on the dynamics of atomic clouds trapped in near-resonant 
optical lattices.
\end{abstract}

%\pacs{32.80.Pj, 34.50.Rk, 42.50.Vk, 03.65.-w}
%}

\maketitle

%\narrowtext

\section{Introduction} 

A periodic polarization or intensity gradient of a laser field 
can create a periodic optical potential structure for atoms.
When atoms move in this structure, they  
may undergo Sisyphus cooling,
and finally 
localize into the optical lattice sites~\cite{Dalibard89,Ungar89,Jessen96,Meacher98,Rolston98,Guidoni99,vanderStraten99}.
After localization, atoms are still able to move around in the lattice.
The optical-pumping-induced motion typically dominates in near-resonant
optical lattices, whereas
in shallow far-off resonant lattices
the quantum mechanical tunneling of atoms between the lattice 
sites may
dominate the atomic motion.
Because of their ability to move between the lattice sites,
two atoms may end up
in the same site and collide.
The purpose of this paper is to present a simple method
to calculate the binary collision rate in a near-resonant
optical lattice.

Cold collisions have been widely studied in magneto-optical
traps~\cite{Weiner99,Suominen96}. This is not the case for optical
lattices where the complications arise due to the position dependent
coupling between the multi-level atoms and the laser field.
So far, to the best of our knowledge, 
there has been only a few experimental
cold collision studies in optical lattices~\cite{Lawall98,Kunugita97}. 
Theoretical studies
of the interactions between atoms in optical lattices
include the
mean-field type
approaches~\cite{Goldstein96,Boisseau96,Guzman98,Menotti99,Menotti99b}, and 
Monte Carlo wave function (MCWF) simulations of binary collisions in red
detuned~\cite{Piilo01,Piilo02b} and blue detuned lattices~\cite{Piilo02}.
However, the mean-field approaches neglect the dynamical nature
of the collisions, and the MCWF simulations for
two atoms colliding in optical lattice require
extremely large computer resources~\cite{Piilo02b}.
It would therefore be useful to find a way to do 
a dynamical study of collisions in near resonant 
optical lattices without the requirement of heavy
computational resources.

We present here a simple method to 
calculate the rate
of binary collisions in near-resonant optical lattices.
The method is based on {\it single-atom} MCWF
simulations~\cite{Dalibard92,Molmer93,Molmer96,Plenio98,Holland94,Holland94b}
and the key idea is to monitor
a quantum flux beyond the average distance $z_{a}$ between the atoms
in a densely populated lattice. 
The accumulation of the
atomic population beyond $z_a$
gives information about the atomic quantum transport
in a lattice and can be monitored
by MCWF simulations for one atom.
Thus, our method avoids the
requirement of large computational  resources of 
the two-atom MCWF collision simulations~\cite{Piilo01,Piilo02b,Piilo02}. 

To illustrate the usefulness of the method we calculate
 the collision rates for various
modulation depths and occupation densities of 
one dimensional optical lattices.
Moreover, we show that 
it is possible to obtain accurate collision rates
for all the densities
from only a few simulations
for a given lattice.
The simulation results show a quadratic 
behaviour of the binary collision rate as a function
of density.

It has been shown before that,
in the parameter regime we use
here, the atomic motion between the lattice sites is to a 
good approximation dominated by the laser-atom interactions 
~\cite{Piilo01,Piilo02b}. The effects of
the interactions between atoms
come into play as soon as the two atoms try to occupy the same
lattice site and collide.
Thus the combination of the method presented here to calculate the
binary collision rate, and the semiclassical methods
accounting the effects of intra-well collisions
~\cite{Weiner99,Suominen96,Suominen98},
can simplify the study of the effects of collisions
on the dynamics of  the atomic cloud trapped in a near-resonant optical
lattice.

We present the lattice structure and our calculation method
of the collision rate in the next section,
the results for various lattice depths and occupation
densities in Section \ref{Results},
and conclude with the discussion in Section \ref{Conclusions}.

\section{Optical lattice and collision rate}\label{Lattice}

\subsection{Sisyphus cooling in optical lattice}

We consider here atoms having ground state angular momentum 
$J_{g}=1/2$ and excited
state angular momentum $J_{e}=3/2$ corresponding to alkali metal elements when
the hyperfine structure is neglected, and use the atomic mass $M$
of $^{133}$Cs. The resonance frequency between 
the states is
$\omega_{0}$ so that $\hbar \omega_{0}=E_e-E_g$, 
where $E_e$ and $E_g$ 
are energies
of the ground and the excited states in zero field. A single atom has 
two ground
state sublevels $|g_{\pm 1/2}>$ and four excited state sublevels 
$|e_{\pm 3/2}>$
and $|e_{\pm 1/2}>$, where  the half--integer subscripts indicate the 
quantum number
$m$ of the angular momentum along the $z$  direction. 

%%% table: used parameters and lattice properties %%%
\begin{table}[t]
\caption[t2]{\label{tab:Parameters}
Used parameters. Rabi frequency $\Omega$, detuning $\delta$,
lattice modulation depth $U_0$, and saturation parameter $s_{0}$.
}
\begin{tabular}{cccccccc}
\hline
       $\displaystyle{\Omega} (\Gamma)$
     & $\displaystyle{\delta} (\Gamma)$     
     & $\displaystyle{U_0} (E_r)$
     & $\displaystyle{s_0}$
 %    & Label 
\\
     \hline
   1.0   & -3.0  & 259 & 0.054  \\
   1.5   & -3.0  & 580 & 0.122   \\  
   1.9   & -3.0  & 936 & 0.195  \\
 \hline  
\end{tabular}
\end{table}
%%% end table %%%

The laser field consists of two counter--propagating beams with 
orthogonal linear
polarizations and with frequency $\omega_L$.
The total field has a polarization
gradient in one dimension and reads (after suitable choices of phases 
of the beams
and origin of the coordinate system)
\begin{equation}
     {\bf E}(z,t)={\cal E}_0 ({\bf e}_x e^{ik_rz} - i {\bf e}_y
     e^{-ik_rz})e^{-i\omega_L t} + c.c.,
     \label{eq:Efield}
\end{equation}
where ${\cal E}_0$ is the amplitude and  $k_r$ the wavenumber.

The intensity of the laser field and the strength of the atom-field coupling 
is described by  the Rabi frequency $\Omega = 2 d {\cal 
E}_0 / \hbar$, where $d$ is  the atomic dipole moment of the strongest
transition between the ground and excited states. The detuning of the 
laser field
from the atomic  resonance is given by $\delta = \omega_L - \omega _{0}$.
As a unit for $\Omega$ and $\delta$ we use the atomic
linewidth $\Gamma$, and express energy in the recoil unit
$E_{r}=(\hbar^{2} k_{r}^{2}) / 2M$.

We keep the detuning fixed, $\delta=-3 \Gamma$, and 
vary the Rabi
frequency $\Omega$, which gives various values for the optical potential
modulation depth
\begin{equation}
    U_0=-\frac{2}{3}\hbar \delta s_0, \label{eq:U_0}
\end{equation}
where $s_0$ is the saturation parameter given by
\begin{equation}
    s_0=\frac{\Omega^2/2}{\delta^2+\Gamma^2/4}. \label{eq:s_0}
\end{equation}
See Table~\ref{tab:Parameters} for used parameters.

The system Hamiltonian after the rotating wave approximation
reads
\begin{equation}
    H_s =  \frac{p^{2}}{2M} - \hbar \delta 
P_{e}
     + V. \label{eq:H0}
\end{equation}
Here $P_{e}$ is the projector operator 
$P_{e} =\sum_{m=-3/2}^{3/2} |e_m
\rangle~ \langle
e_m|$, and the interaction between a atom and the field is
\begin{eqnarray}
    V&=& -i\frac{\hbar\Omega}{\sqrt{2}} \sin(kz)
    \left\{|e_{3/2} \rangle~\langle g_{1/2}| 
%\right. \nonumber \\
%    &&
 + \frac{1}{\sqrt{3}} %\left.
    |e_{1/2} \rangle~\langle g_{-1/2}|\right\} \nonumber \\
    && +\frac{\hbar\Omega}{\sqrt{2}}\cos(kz)
    \left\{|e_{-3/2} \rangle~\langle g_{-1/2}| 
%\right. \nonumber \\
%    &&
  + \frac{1}{\sqrt{3}}
 %\left.
    |e_{-1/2} \rangle~\langle g_{-1/2}|\right\}
\nonumber \\
 && +h.c.
    \label{eq:VAtomLaser}
\end{eqnarray}
The position and time dependent wave function of the system is 
\begin{equation}
       |\psi(z,t)\rangle  = \sum_{j,m}
       \psi_{j,m}(z,t) |j{m}\rangle. \label{eq:Psi}
\end{equation}

The periodic polarization gradient of the laser field 
and the subsequent periodicity of the couplings between the
atomic states are reflected in 
the periodic
light shifts (ac-Stark shifts) of the atomic sublevels creating the optical
lattice structure. When the atomic motion 
occurs in a suitable velocity range, optical 
pumping of the
atom between the ground state sublevels reduces the kinetic energy of the
atom~~\cite{Dalibard89,Ungar89,Jessen96,Meacher98,Rolston98,Guidoni99,vanderStraten99}, and the atom is cooled. 

When the steady state is reached after a certain period of cooling, 
atoms are to a large extent localized into the optical
potential wells. In this study we deal with
near-resonant optical lattices where the laser field is detuned a few
atomic linewidths to the red of the atomic transition. Thus,
after localization atoms may still move around in the lattice
due to the finite extent of the atomic wave packet and 
scattering of photons
which transfers the population between the
various ground state
sublevels and corresponding optical potentials via
the optical pumping mechanism.

Because the atoms are able to move between
the lattice sites,
they may try to occupy the same site and collide.
The purpose of this paper is 
to find a simple way to calculate the binary
collision rate
without the need to perform complicated
two-atom simulations. We use 
the occupation densities
$\rho_{0}$ of the lattice between  $14.2 \%$ and $25.0\%$
corresponding to cases with every seventh and fourth site occupied
respectively. 
Other relevant parameters are presented in Table~\ref{tab:Parameters}.
From the simulation results,
it is possible
to calculate the collision
rates for a wide range of occupation densities 
of the lattice, see Section~\ref{Conclusions}.

\subsection{Calculation of collision rate}

\subsubsection{Monitoring the quantum flux}

\begin{figure}[t!]
\centering
%\vspace{10truemm}
%\psfig{figure=fig1.eps,scale=0.4}
\scalebox{0.4}{\includegraphics{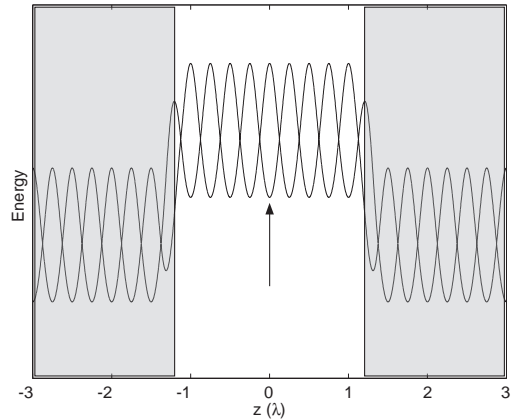}}
\vspace{-15truemm}
\caption[f3]{\label{fig:Opots}
The schematic view of
the optical potentials for the two ground state
Zeeman sublevels. The initial lattice site of the
wave packet is indicated by the arrow and
the modification of the potentials
around the points $\pm z_a$ makes the quantum flux
unidirectional into the accumulation region
(shown as shaded area).
}
\end{figure}

The key idea in our method
is to monitor the quantum
flux beyond the average distance $z_a$ between the
atoms in the one dimensional lattice.
Since the distance between the nearest neighbour
lattice sites is $\lambda/4$, then, e.g., $z_a=\lambda$
corresponds to the occupation density $\rho_0=25\%$
of the lattice. Here $\lambda$ is the wavelength
of the lattice lasers.
We call the area $|z|>z_a$ the accumulation region
(see Fig.~\ref{fig:Opots})
and denote the time dependent fraction 
of the wave packet in this region 
by 
\begin{equation}
%\label{ }
|\psi_a|^2(t)=\int_{|z|>z_a} \psi^\ast(z,t) \psi(z,t) dz.
\end{equation}
The ensemble averaged Monte Carlo (MC) result of $|\psi_a|^2(t)$
gives information about the number of collisions
and the collision rate, as shown below.

To get the value for a collision rate we need to know
how many atoms have travelled into the accumulation
region and in which time.
Thus, we need information about
the quantum flux \textit{into} the accumulation region,
in other words the \textit{cumulative} population in the
accumulation region as a function of time.
This means that we should be
able to make the quantum flux unidirectional
into $|z|>z_a$.
The atoms should therefore be able to arrive into the accumulation
region due to their random motion in the lattice
but should not be allowed to leave.
In this case the accumulation rate of the atomic
population into the accumulation region could give
us the binary collision rate in the lattice.

To make the quantum flux unidirectional into
the accumulation region we add the diagonal 
term
\begin{eqnarray}
\label{eq:Mod}
 H_{m} = \left\{ \begin{array}
%{r@{\quad:\quad}l}
%{r@{\quad}l}
{r@{}l}
 0 :  & |z|< z_a-\lambda/8, \\     
-\alpha: &  |z|> z_a+\lambda/8, \\     
 -\alpha \cos^2\left[ 2 \pi (\frac{z+z_a}{\lambda}+1/8)\right]: & 
 -z_a-\lambda/8\leqslant z \leqslant  \\
 & -z_a+\lambda/8, \\
-\alpha  \sin^2\left[ 2 \pi (\frac{z-z_a}{\lambda}+1/8)\right]: & 
 z_a-\lambda/8\leqslant z \leqslant \\
& z_a+\lambda/8
\end{array}
\right. 
\end{eqnarray}
to the system Hamiltonian, Eq.(\ref{eq:H0}).
We show the total optical potentials
for the two ground state sublevels in Fig.~\ref{fig:Opots}.
The steady state momentum distributions and
the average kinetic energy of the atoms
in the region $|z| < z_{a}$ agree
with the already known results for optical
lattices, see for example Ref.~27.
%\cite{Castin90}
 
When an atom arrives into a lattice site located around $\pm z_{a}$,
corresponding to a collision site,
it is pushed into the accumulation region and is not allowed to climb back
to the region $|z| < z_{a}$. This does not change
the physics for region $|z|< z_{a}$ before the
atom reaches collision site. After arrival
the atom is captured into the accumulation region and
as a consequence it is easy to calculate the cumulative
population as a function
of time.
The trick here is that the drop in energy
can not be too large but has to be large enough:
Sisyphus cooling should still be effective in accumulation region,
preventing the atoms from simply bouncing back from the wall of the grid
and traveling back to $|z|<z_a$;
but the energy drop has to be large enough so
that the probability to climb the energy barrier from
the accumulation region should be small.
We have chosen the values
$\alpha =375,840,1356E_r$ for $U_0=259,580,936 E_r$ respectively.

The total Hamiltonian for the time evolution
of the system now reads
\begin{equation}
H = H_s + H_{m} + H_d,    
\end{equation}
where $H_s$ is the system Hamiltonian 
from Eq.~(\ref{eq:H0}), $H_m$ the optical potential
modification from  Eq.~(\ref{eq:Mod}), and 
$H_d$ includes the non-Hermitian decay part of the MC method
(see below).

%%%%% fig %%%%%
%%% end fig %%%

\subsubsection{Monte Carlo wave function method}

We use the MCWF method (see Refs.~20-23.
%\cite{Dalibard92,Molmer93,Molmer96,Plenio98} 
for details, and, e.g., Refs.~24-26
%\cite{Holland94,Holland94b,Suominen98} 
for applications of the method to the cold collision problems)
to calculate the collision rates.
We have also applied the method recently
to study heating in red-detuned~\cite{Piilo01,Piilo02b}
and optical shielding in blue-detuned lattices~\cite{Piilo02}.

The core idea of the MCWF method is to generate
a large number
of single wave function time evolution histories which
include stochastic  quantum jumps. 
The jumps occur because of the non-Hermitian
part $H_d$ in the system Hamiltonian.
The information about
the decay of the system is included in $H_d$, which shrinks
the norm of the wave function and gives the jump probability
for each time step taken~\cite{jop}. 
The results for the system properties are finally
obtained as an ensemble average
of the generated single wave function histories.

The initial position of the wave packet in our
simulations is given randomly into the lattice well around
the point $z=0$ (see Fig.~\ref{fig:Opots}),
and with zero mean momentum.
Strictly speaking we should use as initial
position and momentum distributions
the steady state distributions for the
used lattice parameters. We emphasize 
that the time
scale to achieve the steady state
is short compared to the population
accumulation time to the accumulation region
and  we have checked that the change of the initial
conditions does not change the simulation results.
Thus, we avoid doing the double effort 
(first calculating the steady state properties
of the system and then the collision rate)
but still get the correct results. If one does 
simulations for higher occupation densities
than used here (which is not actually necessary
since the simulation results calculated
here can be used to obtain the rate curve for
wide range of densities, see Section~\ref{Results}),
then more care should be given to the 
initial conditions of the system.

The number of the generated wave function histories 
per simulation varies from $256$ to $320$
and the total simulation times vary from
$3200\Gamma^{-1}$ to $6400\Gamma^{-1}$.

\subsubsection{Collision rate}

\begin{figure}[t!]
\centering
%\vspace{70truemm}
\scalebox{0.4}{\includegraphics{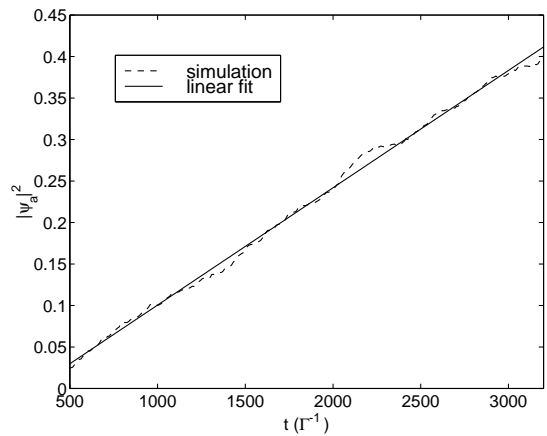}}
\caption[f3]{\label{fig:Cpops}
An example  $(U_0=936E_r, \rho_0=20\%)$
of the accumulation region  population $|\psi_a|^2$ as
a function of time. The binary collision rate $R$
is obtained from the slope of the curve $\beta$ and
the average distance between the atoms $z_a$
as $R=\beta/z_a$, see text. 
}
\end{figure}

We show an example of the ensemble averaged
atomic population in the accumulation region,
$|\psi_a|^2$, as 
a function of time in Fig.~\ref{fig:Cpops}
where we display the simulation result and the linear
fit.

%%%%% fig %%%%%
%%% end fig %%%

Figure~\ref{fig:Cpops} demonstrates a steady flow of
atomic population into the accumulation region.
Thus, by fitting a linear function
to the MC simulation result we can use the
slope of the curve, $\beta$, to calculate
the collision rate $R$ in the lattice.

In the example of Fig.~\ref{fig:Cpops}
$|\psi_a|^2(t=3000\Gamma^{-1}) \approx 0.4$
with the number of histories of 320.
This means that the number of collisions in time
$t=3000\Gamma^{-1}$ in
this specific case was
$0.4$ per $z_a$ or 
$0.4\times320=128$
for a lattice length of $320z_a$.

In general, the collision rate from simulations per unit time and
per unit volume (per unit length in our 
one dimensional lattice) is given by 
\begin{equation}
\label{ }
R=\beta/z_a.
\end{equation}
 The total number
of collisions $N_{tot}$ in the experiment
would
thus be $N_{tot}=R\times t\times  L$ where $t$ is time and
$L$ the length of the one dimensional lattice.

\section{Results}\label{Results}

\begin{figure}[t!]
\centering
%\vspace{70truemm}
\scalebox{0.4}{\includegraphics{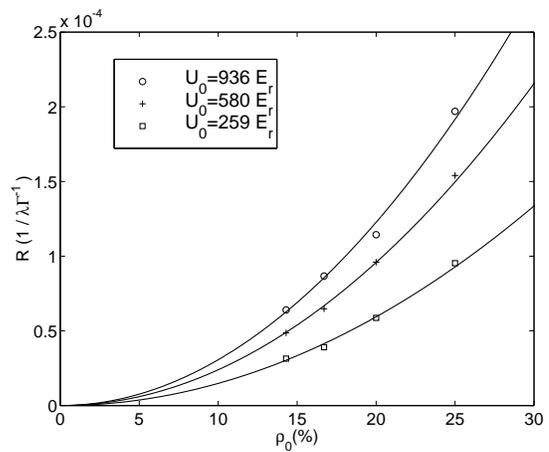}}
\caption[f3]{\label{fig:Crates1}
The binary collision rate $R$
for three different lattice depths $U_0$
as a function of occupation density
$\rho_0$ of the lattice. The points show the simulation
results and the solid lines the quadratic
collision rate curves averaged from the simulation
results for the specific lattice.
}
\end{figure}

We have simulated the binary collision rate $R$
for three different lattice depths $U_0$ with fixed detuning
$\delta=-3\Gamma$ (see Table~\ref{tab:Parameters}), and
for lattice occupation densities 
$\rho_0=14.3\%,16.7\%, 20\%, 25\%$, from every seventh to every fourth
lattice site occupied respectively.
The results are displayed in Fig.~\ref{fig:Crates1}.

The simulation results show a quadratic
behaviour of $R$ with respect to atomic 
density~\cite{quad}.
Thus we can obtain
the whole collision rate curve for all densities of the
specific lattice
by calculating $R$ from only a single density MC result.
By dividing the result for $R$
with the square of the density it is possible get the factor
of quadratic behaviour for all densities.
We note that this might be vulnerable to statistical error
typical for MC simulations.
To improve the statistical accuracy
of the result for all densities
we take the average of 
the quadratic factors obtained for the different
densities.
This makes it possible to obtain the whole collision
rate curve for all the densities of the
specific lattice with good statistical accuracy.
The final results for all the density range obtained
by averaging the various density MC results are drawn
with solid lines in Fig.~\ref{fig:Crates1}.

%%%%% fig %%%%%
%%% end fig %%%

In the parameter regime we use,
the motion between the lattice sites is mainly due to optical
pumping, not because of the tunneling between 
the lattice sites. 
The higher is $U_0$, the larger
is the collision rate.
For fixed detuning the increasing field intensity
means higher $U_0$ and shorter optical
pumping time. The atoms get more mobile
since the increasing excited state population
makes the probability of the  internal state 
changing spontaneous emission event higher.
If the atomic motion between lattice sites was because
of tunneling, then the collision rate would decrease
for tightly bound atoms for larger lattice 
modulation depths.

Moreover, the atomic motion between the lattice sites
is not simply a random walk with a step size
of the lattice constant. 
When the atom changes its internal state and
optical potential well, it is not necessary that the
atom halts its motion already in the nearest
neighbour site. The atom may travel the distance 
of several lattice constants with a single flight,
in fact, for the shallow
optical lattices the atomic motion and diffusion in the lattice
may even be anomalous and include L\'evy walks
~\cite{Bardou94,Marksteiner96}.
For the parameters we use, the diffusion
is normal~\cite{Greenwood97,Visser97}. 

Since the atomic motion between the lattice sites here
is due to optical pumping it is useful to
compare the ratios of calculated collision rates
of the various lattice depths 
to the corresponding ratios of the optical pumping times.
The ratio of the optical pumping times for two different
lattice depths with equal detuning is simply given by the square
of the ratio of Rabi frequencies~\cite{Castin90}.
Here the ratio of the optical pumping times is 
on the same order of magnitude
as the calculated ratios of the collision rates,
but there can be a difference of factor of two.
They do not match exactly
because there is more to atomic motion
between the sites than optical pumping only.
For example the hopping statistics between the sites
may differ for various lattice depths.

\section{Discussion and conclusions}\label{Conclusions}

We have presented a method to calculate the binary
collision rate $R$ in a steady state for an atomic cloud trapped
in a near-resonant optical lattice.
The method is based on MCWF simulations
and the key idea is to monitor
the quantum flux beyond the average distance between 
the atoms. 

We have done the MCWF simulations for  
near-resonant lattice, $\delta=-3\Gamma$, and for lattice
depths $259E_r\leqslant U_0 \leqslant 936E_r$.
From the MCWF results it is possible to calculate the quadratic
collision rate curve for all the densities
of the lattice, within some limitations though (see below).

The advantage of the method is 
the avoiding of the large computational resource requirements
of two-atom simulations.
This is possible because in near-resonant lattices with
large enough modulation depths the atomic motion
between lattice sites is dictated by the field-atom interactions only
~\cite{Piilo01,Piilo02b}.
The interactions between the atoms
come into play only when two atoms
try to occupy the same lattice site and collide
(this is not the case for all parameter ranges,
see Ref.~13).
%\cite{Boisseau96}).
The method is also fast, straightforward, and
fully quantum-mechanical.

In addition of giving quantitative estimate for the number
of collisions in the experiment for a wide density range
of optical lattices, our method
brings out the possibility to combine two simple 
and computationally light methods  
to study collisions in optical lattices.
Namely, the method presented here, combined
with the semiclassical
calculations accounting for the intra-well collision
effects~\cite{Weiner99,Suominen96,Suominen98},
would simplify the study of the effects of the dynamic binary collisions
for the atomic clouds trapped in optical lattices.

We have actually calculated here the collision rate
for the primary collisions in the lattice.
That is: the atomic cloud achieves the steady
state  and it is the steady state properties
which define the collision rate here.
Naturally the dynamics of the cloud
may change when a large fraction
of the atoms collide in short time
for high occupation densities of the lattice.
Moreover, the occupation density may change
if the collided atoms gain enough
kinetic energy to leave the lattice,
or if the lattice is constructed for metastable
atoms~\cite{Lawall98,Kunugita97}, which may ionize and
escape the lattice when 
colliding. This is actually a benefit when using
metastables, since it would
be straightforward, at least in principle, to compare the number of
collisions given by the method presented here and
in the experiment.

We have to neglect the rescattering of photons.
It is practically impossible account the rescattering
effects because of finite amount of available computer
resources.
Despite of this, we think that our study
is useful because of two reasons: a) we can give results for
collision rates in the density region where the rescattering
effects do not appear,  b) because for very
high densities (where rescattering effects should
be accounted in principle) one could still
use the method presented here to study the aspects
of the binary collision effects on the lattice dynamics
of the atomic cloud. This in itself is
very complex problem, and also worth 
studying~\cite{Piilo01,Piilo02b,Piilo02}.

From a Monte Carlo method point of view there
is an interesting feature present here.
Namely, we
first do a few simulations and notice that 
the binary
collision rate has a  
quadratic behaviour with
respect to the atomic density.
As a second step, we obtain the final results 
by taking the average of the various sets
of MC results. We emphasize that this
method
differs from a simple
increase of the number of histories in the MC ensemble
to increase the statistical accuracy.
Thus the benefits here are (in addition to 
the ones mentioned in previous paragraphs)
twofold a) the statistical accuracy of
the results increases b) the result can be
obtained in a wider range than in which the MC simulations
are done. This is a new feature in MCWF simulations to our 
knowledge, at least when MCWF method
is applied to cold collision problems.

\acknowledgments

The author acknowledges financial support from the National Graduate
 School on Modern Optics and
Photonics, discussions with Prof. K.-A.~Suominen, the Finnish Center for
Scientific Computing (CSC) for computing resources, and thanks Matt Mackie
for critical reading of the manuscript.

\newpage

%%% table: used parameters and lattice properties %%%
%\begin{table}[t]
%\caption[t2]{\label{tab:Parameters}
%Used parameters. Rabi frequency $\Omega$, detuning $\delta$,
%lattice modulation depth $U_0$, and saturation parameter $s_{0}$.
%}
%\begin{tabular}{cccccccc}
%\hline
%       $\displaystyle{\Omega} (\Gamma)$
%     & $\displaystyle{\delta} (\Gamma)$     
%     & $\displaystyle{U_0} (E_r)$
%     & $\displaystyle{s_0}$
% %    & Label 
%\\
%     \hline
%   1.0   & -3.0  & 259 & 0.054  \\
%   1.5   & -3.0  & 580 & 0.122   \\  
%   1.9   & -3.0  & 936 & 0.195  \\
% \hline  
%\end{tabular}
%\end{table}
%%% end table %%%

\newpage
\clearpage

\newpage
\clearpage

%%%%% fig %%%%%
%\begin{figure}[t!]
%\centering
%\vspace{70truemm}
%\scalebox{0.4}{\includegraphics{fig2.eps}}
%\includegraphics{fig2.eps}
%\psfig{figure=fig2.eps,scale=0.4}
%\caption[f3]{\label{fig:Cpops}
%An example  $(U_0=936E_r, \rho_0=20\%)$
%of the accumulation region  population $|\psi_a|^2$ as
%a function of time. The binary collision rate $R$
%is obtained from the slope of the curve $\beta$ and
%the average distance between the atoms $z_a$
%as $R=\beta/z_a$, see text. 
%}
%\end{figure}
%%% end fig %%%

\newpage
\clearpage

%%%%% fig %%%%%
%\begin{figure}[t!]
%\centering
%\vspace{70truemm}
%\scalebox{0.4}{\includegraphics{fig3.eps}}
%\includegraphics{fig3.eps}
%\psfig{figure=fig3.eps,scale=0.4}
%\caption[f3]{\label{fig:Crates1}
%The binary collision rate $R$
%for three different lattice depths $U_0$
%as a function of occupation density
%$\rho_0$ of the lattice. The points show the simulation
%results and the solid lines the quadratic
%collision rate curves averaged from the simulation
%results for the specific lattice.
%}
%\end{figure}
%%% end fig %%%


\begin{references}
\bibitem{jop}  See Ref.~18
%\cite{Piilo02b} 
for the
details of the jump operators that are used in
our implementation of the method.
\bibitem{quad} This is in agreement with the result given
in Ref.~10
%\cite{Lawall98}
 where the collisions
are also a measure of atomic transport in a lattice
and are independent of the scattering cross section.
\end{references}
\end{document}